\documentclass[12pt,draftclsnofoot,onecolumn]{IEEEtran}
\usepackage{amsmath,amsfonts}
\usepackage{algorithmic}
\usepackage{array}
\usepackage[caption=false,font=normalsize,labelfont=sf,textfont=sf]{subfig}
\usepackage{textcomp}
\usepackage{color}
\usepackage{stfloats}
\usepackage{url}
\usepackage{verbatim}
\usepackage{graphicx}
\usepackage{epstopdf}
\def\BibTeX{{\rm B\kern-.05em{\sc i\kern-.025em b}\kern-.08em
    T\kern-.1667em\lower.7ex\hbox{E}\kern-.125emX}}
\usepackage{balance}
\begin{document}
\title{Cell-Free XL-MIMO Meets Multi-Agent Reinforcement Learning: Architectures, Challenges, and Future Directions}
\author{{Zhilong~Liu,~\IEEEmembership{Graduate Student Member,~IEEE}, Jiayi~Zhang,~\IEEEmembership{Senior Member,~IEEE}, Ziheng~Liu, Hongyang~Du, Zhe~Wang, Dusit Niyato,~\IEEEmembership{Fellow,~IEEE}, Mohsen Guizani,~\IEEEmembership{Fellow,~IEEE}, and Bo~Ai,~\IEEEmembership{Fellow,~IEEE}}

\thanks {Zhilong Liu, Jiayi Zhang, Ziheng Liu, Zhe Wang, and Bo Ai are with the School of Electronics and Information Engineering and also with the Frontiers Science Center for Smart High-Speed Railway System, Beijing Jiaotong University, China;
Hongyang Du and Dusit Niyato are with the School of Computer Science and Engineering, Nanyang Technological University, Singapore;
Mohsen Guizani is with Mohamed Bin Zayed University of Artificial Intelligence, UAE. (\it{Corresponding author: Jiayi Zhang.)}}
}

\maketitle
\begin{abstract}
Cell-free massive multiple-input multiple-output (mMIMO) and extremely large-scale MIMO (XL-MIMO) are regarded as promising innovations for the forthcoming generation of wireless communication systems. Their significant advantages in augmenting the number of degrees of freedom have garnered considerable interest. In this article, we first review the essential opportunities and challenges induced by XL-MIMO systems. We then propose the enhanced paradigm of cell-free XL-MIMO, which incorporates multi-agent reinforcement learning (MARL) to provide a distributed strategy for tackling the problem of high-dimension signal processing and costly energy consumption. Based on the unique near-field characteristics in XL-MIMO systems, we propose two categories of the low-complexity algorithm design, i.e., antenna selection and power control, to adapt to different cell-free XL-MIMO scenarios and meet the increasing data rate requirement. For inspiration, several critical future research directions pertaining to green cell-free XL-MIMO systems are presented.
\end{abstract}


\section{Introduction}
\IEEEPARstart{T}{he} next generation of wireless communication systems, i.e., the sixth-generation (6G), is expected to deliver unprecedented levels of performance, particularly in digital twins, integrated sensing and communication, and extended reality scenarios. The commercialization of massive multiple-input multiple-output (mMIMO) technology has played a significant role in wireless network development. However, conventional MIMO techniques face limitations in meeting the complex requirements of 6G use cases. In light of this challenge, emerging technologies such as cell-free mMIMO and extremely large-scale MIMO (XL-MIMO) are being proposed to overcome the capacity constraints of conventional MIMO. These advanced technologies are critical to fulfill the massive connectivity and all-round multidimensional access to space, air, ground, and sea, which will enable the Internet of Everything.

As a high-profile technology, the novel cell-free mMIMO holds great promise in meeting the growing demand for increasing network throughput and low-latency transmission. By deploying a large number of geographically distributed access points (APs) connected to a central processing unit (CPU), cell-free mMIMO can effectively address the inter-cell interference that exists in the intrinsic implementation of ``cell-centric'' network \cite{[1],[2]}. Similarly, the promising XL-MIMO technology inherits the prior cellular network with the world-shaking change of base stations (BSs) to adapt the communication variations from far-field to near-field since the massive antennas deployment \cite{[4],[3]}. Moreover, the XL-MIMO can also provide a much stronger beamforming gain as well as harvest abundant degrees of freedom (DoFs) to compensate for the severe path loss in the millimeter-wave and terahertz band communications.

In cell-free mMIMO systems, the data processing procedures can be performed locally using the large-scale fading decoding (LSFD) method \cite{[1]}. This approach is highly effective in relieving the computational load on CPUs. From the perspective of electromagnetic (EM) fields, the addition of antennas in XL-MIMO is a superficial phenomenon. In fact, the significant changes occur in the analysis methods, where the spherical wavefront-based analysis framework replaces the planar wavefront-based one \cite{[17],[9]}. In parallel, the interdisciplinary Electromagnetic Information Theory (EIT) is undergoing a global roll-out research. By integrating cell-free mMIMO and XL-MIMO, namely cell-free XL-MIMO, this prototype will be a forward-looking architecture that can accommodate full scenarios and hot-spot venues to extend the range of near-field communication (NFC), as shown in Fig. 1.

\begin{figure*}[t]
\centering
\includegraphics[scale=0.5]{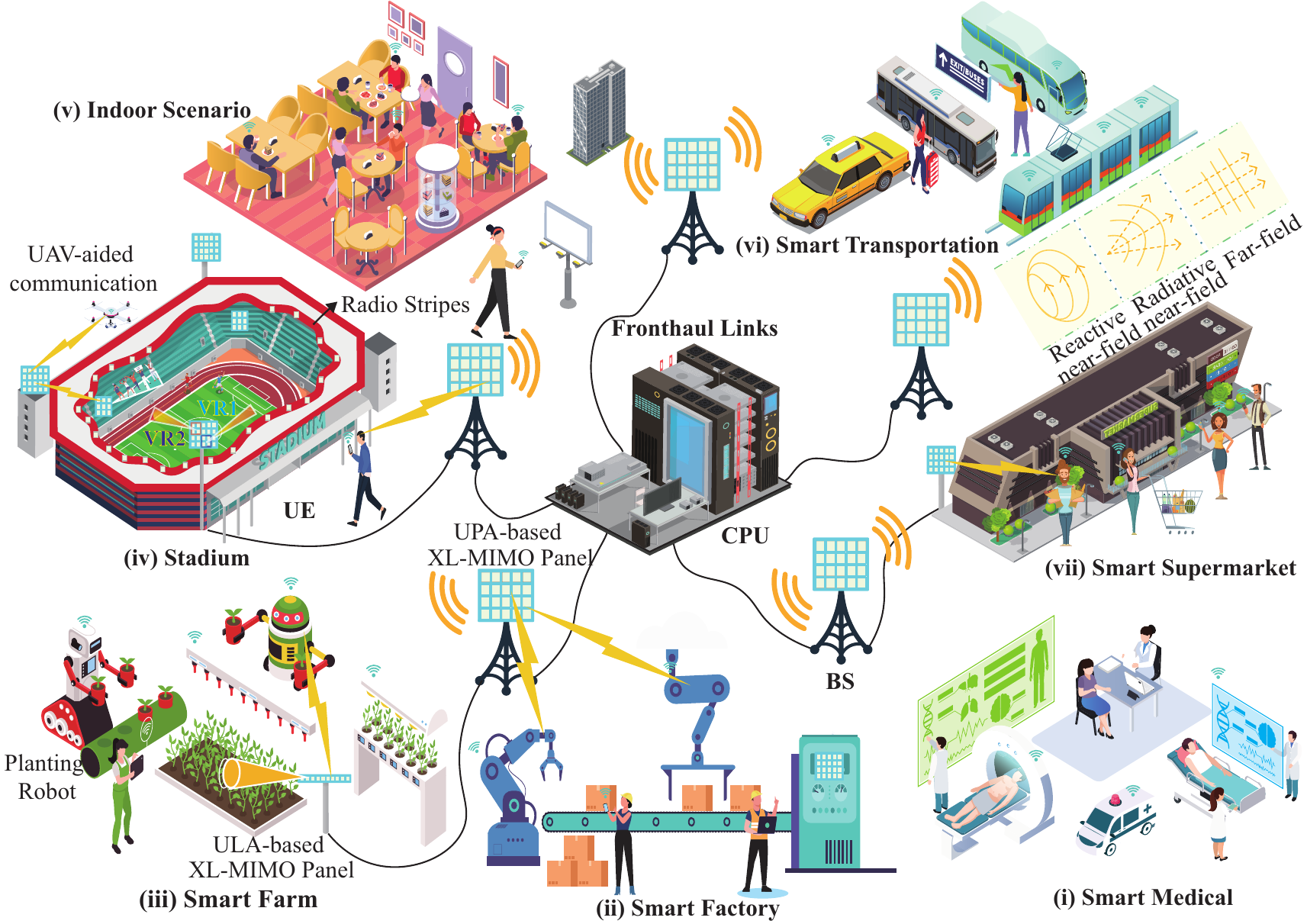}
\caption{System architecture and application scenarios of cell-free XL-MIMO systems around NFC. The BSs are equipped with XL-MIMO panels, and the user equipments are equipped with different numbers of antennas, from single to hundreds. BSs and users are distributed in the service area. The BSs are connected to a CPU with a high computation processing ability via fronthaul links. The communication regions are divided into reactive near-field, radiative near-field, and far-field \cite{[3]}. In XL-MIMO systems, the communication focuses on the radiative near-field. The boundary between radiative near-field and far-field is decided by the Rayleigh distance \cite{[19],[10]}, and visibility regions (VRs) induced by the non-stationary channel are illustrated since sheltering from different buildings and obstacles. In 2019, Ericsson proposed radio stripes, the prototype of cell-free mMIMO systems, which is an ideal deployment solution for outdoor and indoor areas such as shopping malls, stadiums, smart factories, and other scenarios \cite{[1]}.}
\end{figure*}

To reduce the overall system computational complexity and energy consumption, low-complexity baseband signal processing algorithms are in demand. Multi-agent reinforcement learning (MARL) has been widely used for decision-making in large-scale network scenarios \cite{[5],[7],MADDPG}, e.g., unmanned aerial vehicles (UAVs), swarm intelligence, and traffic scheduling. We hasten to say that the algorithms are proficient to improve spectral efficiency (SE), enhance interference management in XL-MIMO systems, and to increase coverage, improve user fairness, and achieve distributed resource allocation in cell-free mMIMO systems. In multi-agent systems, interactions between intelligent agents and environments drive the achievement of goals. In particular, RL algorithms become almost indispensable tools for exploring complex dynamic scenarios, which can effectively reduce the overall power consumption. Notable advances include the development of low-complexity RL-based power control algorithms that can be scaled to XL-MIMO systems and the exploration of hybrid analog-digital precoding schemes that can considerably enhance the energy efficiency (EE).

Motivated by the aforementioned works, we investigate the cell-free XL-MIMO systems with MARL techniques. The main contributions are summarized as follows:

\begin{itemize}
\item[$\bullet$] We introduce new NFC characteristics, basic system scheme, and application scenarios of cell-free XL-MIMO systems. More important, we comprehensively introduce the crucial challenges of power consumption, computational complexity, and user mobility.
\end{itemize}

\begin{itemize}
\item[$\bullet$] We investigate three technical frameworks, e.g., fully decentralized, fully centralized, and centralized training and decentralized execution (CTDE), algorithm categories, and applications of MARL methods in existing literature, as shown in Fig. 2.
\end{itemize}

\begin{itemize}
\item[$\bullet$] To strive for the undiscovered performance, we focus on two critical methods, i.e., antenna selection (AS) and power control, to reduce power consumption and improve SE with MARL methods. Numerical results are given to illustrate the ability to improve SE and EE. Finally, the article concludes by discussing open problems toward uncovering the potential of cell-free XL-MIMO systems.
\end{itemize}

\section{Opportunities and Challenges of XL-MIMO communication systems}
\par In this section, we focus on the newly discovered EM wave transmission characteristics in the NFC domain. Through the unique near-field properties uncovered by the XL-MIMO, such as the spherical wave model (SWM), spatial non-stationary effect, and effective DoF (EDoF), they can be well designed to enhance the communication performance. In addition, the power consumption, computational complexity, and mobility problems present us with new challenges.

\subsection{New Opportunities}
\begin{itemize}
\item[$\bullet$] \emph{Spherical Wave Model}
\end{itemize}

SWM is a mathematical tool used to describe the behavior of EM waves in three-dimensional space \cite{[3]}. An accurate SWM is essential for excavating the capacity upper bound of XL-MIMO systems as it facilitates the efficient processing and manipulation of EM waves, thereby improving signal quality and enhancing data throughput. In previous research, channel models mainly focused on the basic assumption of Rayleigh or Rician fading channels. However, once the communication distance is shorter than the Rayleigh distance, e.g., for an XL-MIMO panel with a diagonal of 10 m at 3 GHz, the boundary is up to 2 km, the communication domain focuses on the near-field rather than the far-field. Therefore, the existing channel models used to analyze the conventional MIMO systems are not suitable for XL-MIMO systems as the NFC dominates \cite{[19]}.

Furthermore, the integration of massive antennas can make it difficult to obtain accurate channel state information (CSI) in XL-MIMO systems. Regarding the near-field effects, the channel should be properly modelled to ensure accuracy in the near-field under the spherical wavefront assumption. Based on EIT, the exploration of SWM has revolutionized for enabling high-speed data transmission, broadening coverage and improving user experience \cite{[18]}.

\begin{itemize}
\item[$\bullet$] \emph{Spatial Non-stationary Effect}
\end{itemize}

In XL-MIMO systems, the spatial non-stationary effect arises because only partial antennas in BSs can receive spherical EM waves from specific UEs propagated by different scatters. This can lead to fluctuations in the channel gain, phase, and delay over time \cite{[18]}. Similarly, each UE can only observe a subset of the antenna array, which is called the visibility region (VR), as shown in Fig. 1. As a result, the channel capacity and quality vary significantly, and the traditional channel estimation (CE) and equalization techniques may not be effective at mitigating the effects of non-stationary channels. Thus, effectively exploiting this peculiarity would be a tutorial for green communication systems, as a cost-effective way to reduce the computational complexity for crowded scenarios.

\begin{itemize}
\item[$\bullet$] \emph{Effective Degree of Freedom}
\end{itemize}

An important parameter characterizing the performance of XL-MIMO systems is the EDoF, referring to the number of significant electromagnetic modes. It represents the potential capacity of a MIMO system to spatially multiplex multiple data streams. The EDoF considers the effects of various factors, e.g., channel correlations, signal-to-noise (SNR) ratios, and interference. However, increasing the number of antennas may not always improve the EDoF, as it may increase channel correlation, interference between different data streams, and energy consumption, all of which degrade the system performance \cite{[3]}. Therefore, to achieve a high EDoF in practical XL-MIMO systems, appropriate antenna numbers and configurations should be chosen based on specific wireless channels and system requirements, e.g., the maximum EDoF is around 1600 for a 2.25 m $\times$ 2.25 m panel size at 0.1 m wavelength to satisfy the hot-spot scenarios.

\subsection{New Challenges}

\begin{itemize}
\item[$\bullet$] \emph{Power Consumption}
\end{itemize}

Although the XL-MIMO technology can effectively improve the speed and reliability of the signal transmission, implementing enormous sub-processing units in the XL-MIMO transceiver can result in high hardware cost and power consumption. Reducing the power consumption in the XL-MIMO is essential to ensure the energy-efficient, cost-effective, and sustainable while maintaining its high performance capabilities \cite{[16]}. Currently, existing methods solving the above challenges mainly focus on traditional methods, e.g., heuristic fractional power control laws and deep learning-based power control methods. It is necessary to balance the system performance and power consumption factors considering the characteristics in practical NFC scenarios.
\begin{itemize}
\item[$\bullet$] \emph{Computation Complexity}
\end{itemize}

Complex computations significantly increase latency in wireless communication systems with limited computing power. Distributed signal processing methods are trends to deal with the high-dimension computation. By delegating the processing authorities to the local processing unit (LPU), XL-MIMO can reduce latency and specialized processors or additional memory requirements. In general, the antenna selection technique involves selecting the appropriate subset of antennas from the antenna array, which can help minimize the computational burden of signal processing and power consumption, as well as improve SNR \cite{[12]}.

\begin{itemize}
\item[$\bullet$] \emph{User Mobility}
\end{itemize}

In XL-MIMO systems, user mobility leads to time-varying channel conditions. As users move within the coverage area, channel characteristics, such as path loss, fading, and interference, change dynamically. Moreover, the movement of UEs can switch the propagation mode between the near-field and far-field, and thus, the channel estimation and codebook design in the hybrid-field should be re-examined. Additionally, user mobility necessitates dynamic adaptation of transmission strategies, handover management, user scheduling, and mobility prediction. By considering these factors, XL-MIMO systems can be effectively optimized to maintain reliable connectivity.

With these aspects, the XL-MIMO can be seen as an extended version of the conventional MIMO, which involves more than just increasing the number of antennas deployed from 64 antennas to 256 antennas. From an environmentally-friendly perspective, by optimizing the transceiver power and adopting appropriate distributed processing algorithms, XL-MIMO systems can achieve a superior performance while minimizing the energy consumption and reducing the carbon footprint of wireless communication systems.

\section{System Architecture of Multi-agent Cell-Free XL-MIMO}

With the increasing computational dimension and time-varying configurations and parameters, the traditional optimization methods do not work well with XL-MIMO systems. We have to seek a better solution to resolve it. In what follows, by integrating the advantages of the cell-free mMIMO and MARL methods, we propose a novel cell-free XL-MIMO system with the MARL optimization scheme to further improve the performance of cell-free XL-MIMO systems.
\subsection{Multi-agent Reinforcement Learning}
\begin{figure*}
\centering
\includegraphics[scale=0.3]{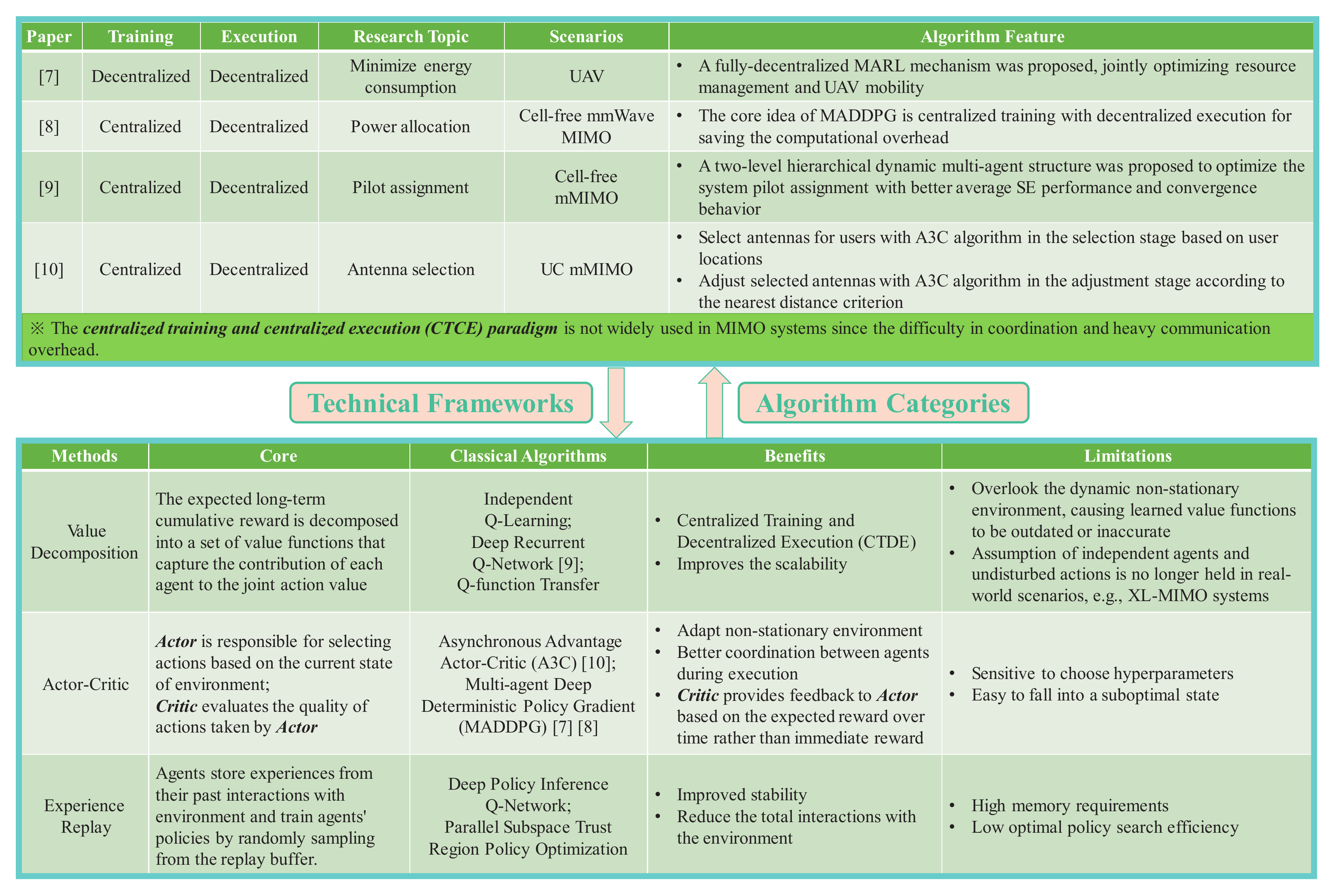}
\caption{Summary of mainstream MARL technical frameworks, including Decentralized Training and Decentralized Execution (DTDE) and Centralized Training and Decentralized Execution (CTDE), algorithm categories, and applications in the existing literatures [7]-[10].}
\end{figure*}

\begin{figure*}[t]
\centering
\includegraphics[scale=0.6]{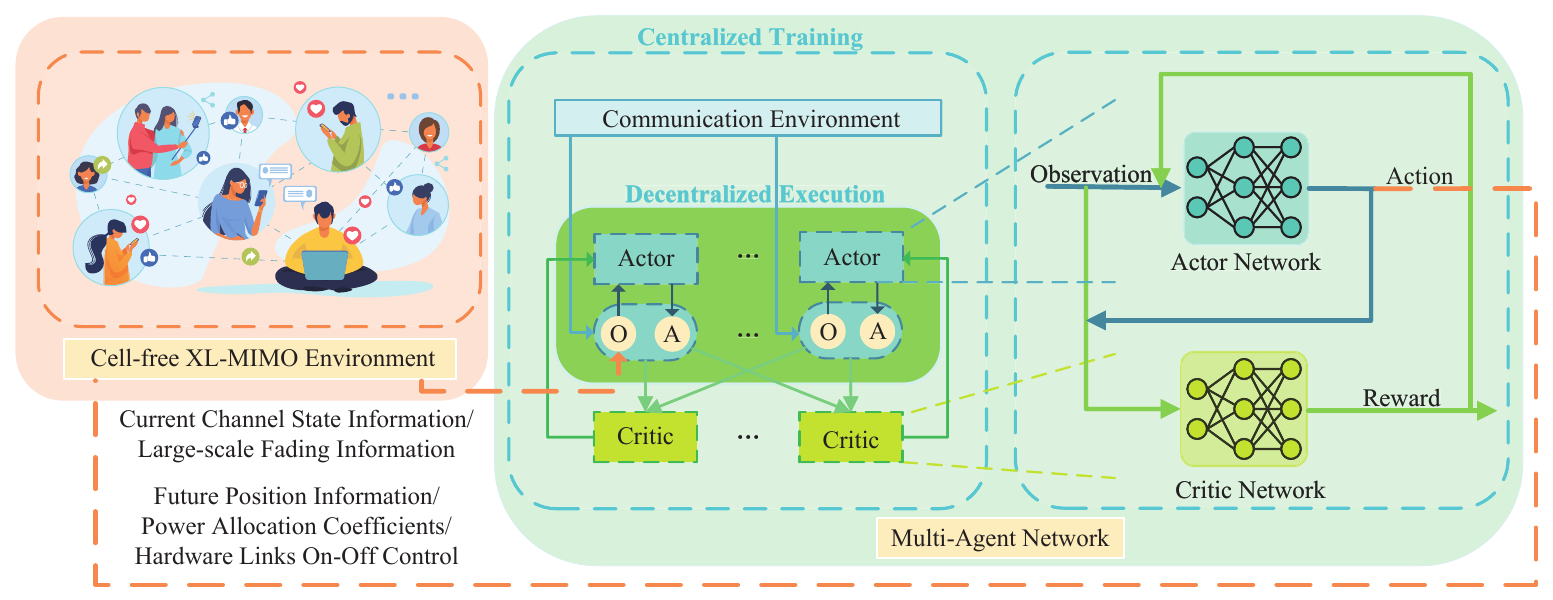}
\caption{The basic scheme of CTDE-based MADDPG algorithm interacting with cell-free XL-MIMO systems.}
\end{figure*}

MARL, a subfield of artificial intelligence, has been widely used in real-world scenarios focusing on the interaction with the environment and multiple agents. Extending from a single-agent domain to a multi-agent environment, this method arises from the need to develop intelligent systems that can interact with other intelligent agents in complex and dynamic environments. It combines the principles of RL, game theory, and multi-agent systems to enable agents to learn how to interact with other agents and the environment to achieve their goals \cite{[5]}. The main idea behind MARL is to model the behavior of a group of agents that can cooperate, compete, and even negotiate. More intuitively, different training schemes, e.g., fully decentralized, fully centralized, and centralized training and decentralized execution (CTDE), are considered as promising paradigms to adapt to different environments. Furthermore, the existing MARL algorithms can be divided into three categories.
\begin{itemize}
\item[$\bullet$] \emph{Value Decomposition}
\end{itemize}

Value decomposition (VD) based algorithms are usually based on value functions, i.e., \emph{Deep Recurrent Q-Network}, to decompose value functions into local value functions for agents, so as to deal with the interaction between multiple agents. This type of algorithm usually combines the actions and states of multiple agents as global states, and then uses single-agent algorithms such as \emph{Q-learning} to learn local value functions.

\begin{itemize}
\item[$\bullet$] \emph{Actor-Critic}
\end{itemize}

Actor-Critic (AC) based algorithms combine \emph{value functions} and \emph{strategy functions} with two networks, \emph{\textbf{Actor}} and \emph{\textbf{Critic}}, where the \emph{Actor network} learns the strategy and the \emph{Critic network} evaluates the value of the action and updates the actor's policy. Examples of AC-based methods include Asynchronous Advantage Actor-Critic (A3C) \cite{[8]} and Multi-agent Deep Deterministic Policy Gradient (MADDPG) \cite{MADDPG}. To illustrate, MADDPG follows the CTDE paradigm, where the additional information has been gathered in \emph{Critic networks} to faciliate the training process while \emph{Actor networks} take actions based on their own local observations.

\begin{itemize}
\item[$\bullet$] \emph{Experience Replay}
\end{itemize}

Experience replay (ER) based algorithms typically use experience replay caches to store past experiences and randomly sample them for training. This approach speeds up training by making more efficient use of data, and is usually applied experiential playback to single-agent algorithms, i.e., \emph{Deep Policy Inference Q-Network}. However, in multi-agent scenarios, the implementation of experience reply is more complicated, and the interaction between the multi-agent needs to be considered.

As shown in Fig. 2, these three categories have been widely used in communication scenarios for resource allocation. While the centralized learning method is advantageous for global assessment with unified decision-making, distributed learning using the MARL methods is more feasible for local processing, which is beneficial for real-time processing.

In multi-agent environments, agents' actions affect the state of the environment, and each agent must learn a policy that not only maximizes its rewards but also takes into account the actions of other agents. The MADDPG algorithm extends the popular DDPG algorithm by introducing a centralized \emph{Critic network} that can observe the joint actions of all agents and provide feedback to each agent's policy network, as shown in Fig. 3. In turn, the \emph{Actor network} learns to optimize their policies, taking into account the feedback from the \emph{Critic network} and the observations of other agents.

In the signal processing phase of the XL-MIMO, high-dimensional matrix operations, and time-sensitive actions are critical to achieve the optimal system performance. Therefore, traditional data processing schemes no longer meet the requirements of cell-free XL-MIMO systems. As such, we have to concentrate on local processing or distributed signal processing to reduce the load on the fronthaul links. For example, we can apply the MARL methods to approach the SE or EE maximum by defining a Markov decision process that includes states, actions, and rewards \cite{[5]}. The agents interact with the environment in the current state and move to the next state. Then, the next state is sent to the agent, which decides to take an action against the environment. The environment then sends the next state and reward to the agent.

\subsection{System Architecture of Multi-Agent Cell-Free XL-MIMO}

\par In conventional massive MIMO systems, centralized processing methods lack the ability to parallelize operations. Furthermore, scaling up the dimensions of the array proves to be an arduous feat owing to the significant amount of interconnections and overwhelming burden placed on the central node. Therefore, various decentralized techniques have been proposed. Among them is the cell-free architecture, which aims at eliminating cell boundaries and focusing on user-centric communication \cite{[1]}, providing more flexible transmission/reception of UEs. To adapt to the requirements of distributed architectures, we propose a modified embodiment of distributed XL-MIMO that exploits the advantages of cell-free mMIMO systems while considering multi-agent systems simultaneously.

As shown in Fig. 1, a distributed-processing XL-MIMO system architecture drawing on the merits of cell-free mMIMO is illustrated. The so-called LSFD method can be used to detect the signals using maximum ratio combining or minimum mean squared error combining \cite{[1],[16]}. For each BS equipped with XL-MIMO panels, it completes the signal processing as well as the channel estimation with all the CSI. All processed signals are then transmitted to the CPU via fronthaul links. In cell-free XL-MIMO systems, there are multiple antennas at the transmitter and receiver sides, and a large number of UEs communicating simultaneously. The communication and resource allocation between these antennas and users can be optimized using MARL, a technique that allows agents to learn how to behave in an environment by interacting with it and receiving feedback in the form of rewards.

Using MARL, agents, i.e., UEs, BSs, and even antennas, can learn to allocate physical layer resources and optimize the transmission strategy further. They interact with the system environment with their CSI and location for acquiring the future decision until the SE or EE maximum is reached. Besides, the MARL-based approach can adapt to dynamic changes in the environment, such as UE mobility and time-varying channels.

\section{Directions and Solutions of Multi-agent Cell-free XL-MIMO systems}

\par Multi-antenna technology has been widely recognized as an effective means of improving SE with diversity gain and multiplexing gain. However, to achieve more performance gains, the computational complexity of cell-free XL-MIMO systems increases rapidly with the number of antennas and grievous interference causes signal quality degradation.

Having introduced the new opportunities, in this section, we provide a new look to solve the urgent challenges with MARL methods, e.g., AS and power control.

\subsection{Challenge 1: Antenna Selection}
\begin{figure}[t]
\centering
\includegraphics[scale=0.4]{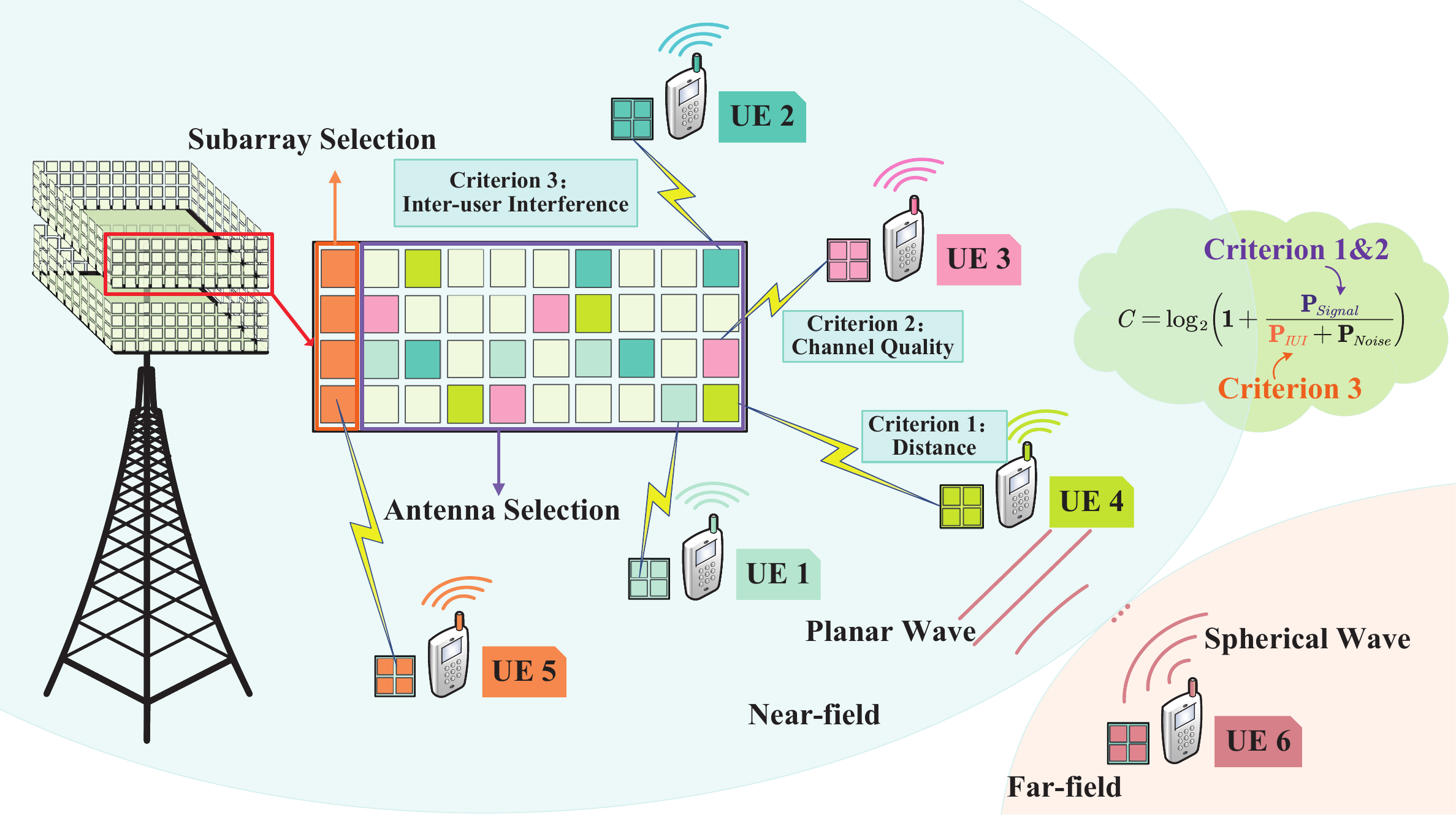}
\caption{The antenna selection of the BS that is equipped with a planar XL-MIMO. A BS simultaneously serves four UEs with different antennas, and different antennas serve different UEs without reuse. The different AS strategies depend on various criteria, e.g., maximizing the received signal power (criterion 1 and criterion 2), minimizing the inter-user interference (criterion 3) and so on.}
\end{figure}

\emph{1) MARL-empowering Antenna Selection}

In cell-free XL-MIMO systems, it is necessary to explore effective AS techniques to reduce the number of antennas used in work patterns, enhance performance, and minimize complexity, especially in energy-constrained environments \cite{[12]}. Not all antennas serve uplink or downlink UEs simultaneously, making it possible to reduce the number of radio-frequency (RF) links and signal processing units to lower hardware cost and power consumption. As the number of antennas tends to be enormous, the circuit cost and computational complexity of conventional methods based on fully-digital receive arrays will increase dramatically.

AS provides a low hardware-complexity mentality for exploiting the spatial-diversity benefits of multiple antenna technology with solely partial antennas activated to serve different UEs and can be considered at both transmitters and receivers in cell-free XL-MIMO systems. The basic idea of AS is to choose the optimal subset of antennas from the available antennas in the whole antenna array, based on some selection criteria \cite{[12]}, as shown in Fig. 4. In a cell-free XL-MIMO, AS can be achieved either statically or dynamically. In static AS, a fixed set of antennas is selected that remains unchanged during transmission, while in dynamic AS, the optimal set of antennas is determined based on the channel conditions at each transmission.

\begin{figure}[t]
\centering
\includegraphics[scale=1]{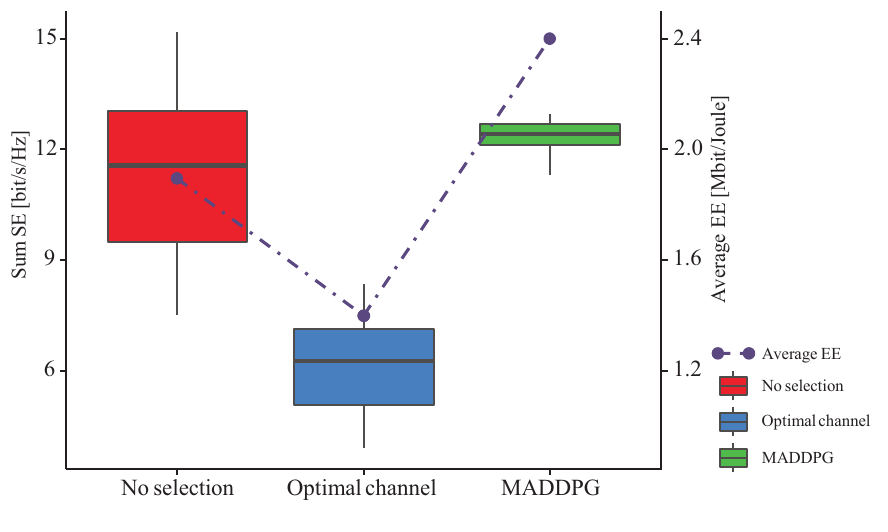}
\caption{Boxplot of sum uplink SE and average EE of a cell-free XL-MIMO system under three circumstances: no selection (with all antennas of XL-MIMO panel activated), optimal channel selection based on the large-scale fading coefficients, and MADDPG-based selection. In the simulation, we consider the LSFD architecture \cite{[16]} with single BS and multiple UEs. All BS and UEs are equipped with XL-MIMO panels. The BS with $N_{r}$ = 81 antennas serving $K$ = 6 UEs with $N_{s}$ = 9 antennas each simultaneously within a square of size 1000 m $\times$ 1000 m. Additionally, the data transmission power $p$ = 200 mW.}

\end{figure}

In Fig. 5, we draw the boxplot of the sum SE and average EE under different AS strategies. %By defining the action based on the channel state and
Each UEs' antenna is regarded as an individual agent to select different BS antenna for achieving the maximum SE. For a fair comparison, we assume the case without AS as a benchmark. The traditional optimal channel selection based on LSF coefficients decreases the system performance by sacrificing DoFs. However, with the introduction of ``multi-agent", each antenna can dynamically adjust the selected antennas . It is noteworthy that the MADDPG algorithm can effectively improve the SE of poor quality UEs and nearly achieve a 26\% EE improvement compared with the benchmark.

\emph{2) Future Research Directions}
$\bullet$ \emph{Hardware Design:} To overcome computationally complex bottlenecks, one promising solution is to partition the uniform planar array (UPA) or uniform linear array (ULA)-based XL-MIMO into \emph{subarrays-disjoint units} with partial-connected structure and individual processing units. Instead of connecting all the antennas, only a subset of antennas is interconnected, allowing antennas to be connected in a flexible and scalable manner.

$\bullet$ \emph{Subarray Selection:} Apart from AS, \emph{subarray selection} is worth investigating with fixed or adjustable format depending on whether they correspond to separate hardware entities or software-defined logical connections between different antenna elements, as shown in Fig. 4. The use of subarrays enables more efficient and distributed processing, enabling the system to handle larger and more complex data sets without compromising on performance and accuracy.

$\bullet$ \emph{Non-stationary Perspective:}
One approach to achieving AS in non-stationary channels is to use multiple antennas in combination with channel estimation and equalization techniques, such as space-time coding and beamforming. These techniques can help mitigate the effects of non-stationary channels by using multiple antennas to create a more robust signal and adapting the transmit and receive strategies to the changing channel conditions.

\subsection{Challenge 2: Power Control Design}
\emph{1) Existing Power Control Method}
\par Apart from the AS, designing an effective power allocation algorithm is another open challenge for reducing power consumption in cell-free XL-MIMO systems. With limited communication resources, the dynamic power allocation is worth optimizing based on the real-time channel information. The existing power control methods solving the inter-user interference are focused on the following optimization objectives: max-min, max-product, and max-sum. Traditional power control methods, such as linear optimization techniques, have limitations in large-scale MIMO systems due to the increased complexity and static configuration. Though the non-convex problem can be easily solved using supervised learning-based methods or centralized mechanisms, the prior optimal output data is challenging to obtain in large-scale networks.
\par With the benefits of massive antennas, the cell-free XL-MIMO poses new challenges for power optimization. Affected by the near-field propagation and spherical wavefront, different parts of the extremely large array encounter different signal strengths. Besides, certain antennas may have minimal impact on the overall system performance due to the non-stationarities and VRs. These lead to the activation of power-intensive RF links for these antennas becoming burdensome and significantly reducing the total EE of systems. In this case, existing algorithms are not always able to harvest the global optimal solution, especially when dealing with high-dimensional matrix operations. To overcome these limitations, MARL algorithms have been applied to deal with power control in cell-free XL-MIMO systems.
\emph{2) Proposed MARL-based Power Control Method}

RL algorithms enable real-time optimization of power control decisions based on the current state of the system, including channel conditions and signal quality. The basic idea behind using MARL algorithms for power control in large-scale MIMO systems is to model each BS or antenna as an individual agent and to optimize the joint behavior of all agents using RL techniques. This allows for a more flexible and data-driven power control solution compared to traditional methods.

To achieve power control in large-scale MIMO systems using MARL algorithms, the following steps can be taken:

$\bullet$ \textbf{\emph{Select individual agent:}} Each antenna, BS, or UE can be modeled as an independent agent, with its unique state, action, and reward, depending on the uplink or downlink transmission. The state of the agent should represent the current channel conditions and interference, while the action should represent the transmit power of the antenna.

$\bullet$ \textbf{\emph{Define reward function:}} The reward function should reflect the performance objective of the power control algorithm, such as maximizing SE or minimizing interference.

$\bullet$ \textbf{\emph{Train MARL algorithm:}} MARL algorithms should be trained using the defined reward function and the modelled agents. The training process involves multiple iterations of the agents taking actions, observing the results, and updating their policies based on the reward received.

$\bullet$ \textbf{\emph{Implement power control algorithm:}} Once the training process is complete, the power control algorithm can be implemented in large-scale MIMO systems. The algorithm will use the learned policies of the agents to determine the optimal transmit power of each antenna.

$\bullet$ \textbf{\emph{Evaluate performance:}} The performance of power control algorithms should be evaluated in a realistic simulation or test environment to ensure their effectiveness in cell-free XL-MIMO systems.

\begin{figure}[t]
\centering
\includegraphics[scale=1]{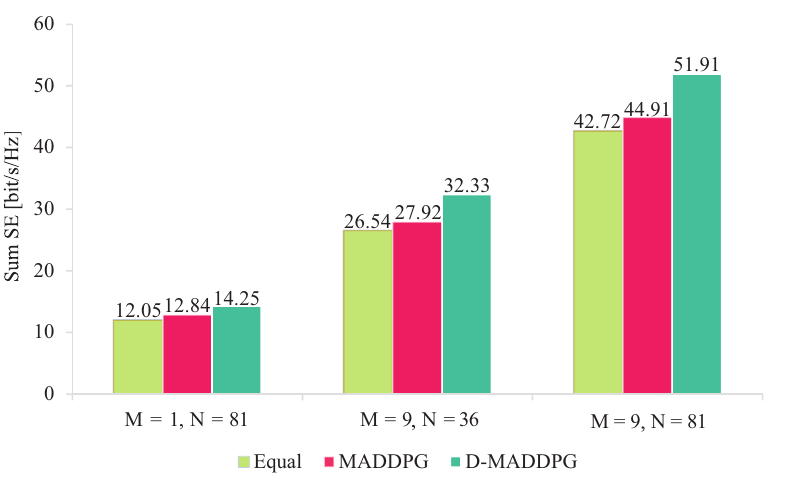}
\caption{Sum uplink SE of cell-free XL-MIMO systems with different power control algorithms: the equal power method, MADDPG strategy, and D-MADDPG under different BS number $M$ and XL-MIMO antennas number $N$. In the D-MADDPG architecture, we model the power control problem as the MARL framework with two layers. In the first layer, each agent corresponds to a BS in the cell-free XL-MIMO system, and the objective is to optimize its transmit power level to maximize system performance while taking into account interference from other agents. Then, with the constraint of BS power obtained in the first layer, the second layer is responsible for the allocation of each antenna of XL-MIMO panels. The simulation parameters are the same with Fig. 5.}
\end{figure}

The application of MARL algorithms for power control in large-scale MIMO systems is a growing area of research, and recent studies have demonstrated the potential of these algorithms for improving the performance and efficiency of MIMO systems \cite{[16]}. Based on existing MARL methods, we successfully apply the MADDPG algorithm to solve power control problems for better performance. In addition, we introduce a double-layer power control architecture called D-MADDPG that is based on LSF coefficients between antennas. This architecture differs from the conventional single-layer architecture, which considers all antennas subjected to an agent as a whole, and it demonstrates a notable advantage in increasing the sum SE, as shown in Fig. 6.

\emph{3) Future Research Directions}

$\bullet$ \emph{Precoding Design:}
The hybrid precoding is promising to relieve the pressure of excessive power consumption by decomposing the high-dimensional full-digital precoder into the realization of an analog beamformer and digital precoder. Through effective precoding design, the RF links and power costs can be significantly reduced. Additionally, advanced precoding designs can mitigate the beam split effect that severely degrades the achievable rate degradation.

$\bullet$ \emph{Partial Interaction Design:}
Designing a distributed MARL algorithm with partial-interaction architecture is a promising way to lessen the quantity of network training and information interaction. Partial-interaction allows agents to selectively select appropriate agents for interaction based on distance, service relationship, and other factors, rather than interacting with all agents in cell-free XL-MIMO systems, which is more practical for scalable networks.

$\bullet$ \emph{Jointly Optimization Design:}
The jointly optimized design of AS and power control is promising to enhance the robustness of the system, eliminating the need for separate optimization. And the power allocation can be re-examined with appropriate antennas selected from XL-MIMO systems.

Based on the above discussion, the design of AS and power control based on real-time interactions with the MADDPG method achieves a higher performance gain in the near-field. Accordingly, such effective MARL methods can be extended to other resource allocation schemes.

\section{Future Research Directions}
\subsection{Hybrid-Field Channel Estimation}
To obtain accurate CSI, CE is a key challenge because the near-field angle-domain channel is not sparse. Faced with huge data streams, lightweight CE methods with reduced computational complexity, fast convergence, and exhaustive channel feature capture are essential to adapt to the near-field characteristics and non-stationary channels. Furthermore, the accurate models based on the spherical wavefronts even the hybrid spherical- and planar-wavefronts, which capture more channel details are essential to reduce the bit error rate due to the user mobility.
\subsection{Hybrid-Field Beamforming}
First, for the near-field beam training, the array response vector of near-field channels is not only related to the angle but also the distance, resulting in a high-dimension codebook set. Thus, a polar-domain codebook should be utilized instead of a discrete Fourier transform codebook to capture the information on the channel paths. Secondly, the near-field beam split effect occurs when the transmitting antennas are placed close to each other and the distance between the antennas is comparable to the wavelength of the signal being transmitted. In such cases, the transmitted signal may split into multiple near-field beams that interfere with each other. Thirdly, it is a hybrid-field joint design optimization for solving the switch from far-field beamsteering to near-field beamfocusing, and vice versa.
\subsection{RIS-aided Cell-free XL-MIMO}
With the ability to dynamically reconfigure the electromagnetic environment, reconfigurable intelligent surface (RIS) can improve channel quality and overcome the limitations of the propagation environment. In the future, the evolution of RIS will perhaps develop towards extremely large-scale RIS (XL-RIS) for the future 6G wireless communications, which makes beam training complicated and data throughput exploded. Additionally, since the RIS is deployed in the near-field of XL-MIMO, the RIS codebook should be well-designed considering the NFC characteristics.
\subsection{Green Communications}
To achieve green communications, next-generation communication systems propose sustainable, energy-efficient, and energy-aware requirements. Low-resolution devices, e.g., analog-to-digital converters (ADCs), are the trend to cope with the great expense of cell-free XL-MIMO systems. On the one hand, hardware impairments still confuse signal processing, especially when the dimension is gigantic. Accordingly, the fruitful compensation algorithm design is necessary to approach the optimum. On the other hand, the simultaneous wireless information and power transfer technology should focus on elaborate near-field beamforming design to achieve a higher performance.
\section{Conclusion}
\par In this article, the fundamental opportunities in the near-field communication of cell-free XL-MIMO systems and open challenges have been discussed in terms of SWM, spatial non-stationary effect, EDoF, power consumption, and computational complexity, respectively. In particular, we investigated the existing MARL categories and proposed the basic scheme of promising cell-free XL-MIMO systems using MARL methods. Then, we started with two existing challenges namely AS and power control. Accordingly, we successfully applied MADDPG algorithms to solve them. Finally, we pointed out the critical and promising future research directions, which are hybrid-field CE, hybrid-field beamforming, RIS-aided cell-free XL-MIMO architecture, and green communications.
\section{Acknowledgments}
This research is supported in part by the Fundamental Research Funds for the Central Universities under Grant No. 2023YJS001, in part by National Key R\&D Program of China under Grant 2020YFB1807201, in part by National Natural Science Foundation of China under Grant 62221001, in part by Natural Science Foundation of Jiangsu Province, Major Project under Grant BK20212002, in part by the Fundamental Research Funds for the Central Universities under Grant 2022JBQY004, in part by ZTE Industry-University-Institute Cooperation Funds under Grant No. HC-CN-20221202003, in part by the National Research Foundation, Singapore, and Infocomm Media Development Authority under its Future Communications Research \& Development Programme, DSO National Laboratories under the AI Singapore Programme (AISG Award No: AISG2-RP-2020-019), and MOE Tier 1 (RG87/22).
\section*{Biographies}
\textbf{Zhilong Liu} received the B.S. degree from the School of Information and Control Engineering, Qingdao University of Technology, Qingdao, China, in 2022. He is currently pursuing the Ph.D. degree with Beijing Jiaotong University, Beijing, China. His research interests include massive MIMO systems, signal processing, reinforcement learning, and performance analysis of wireless systems.
\vspace{0.1cm}

\textbf{Jiayi Zhang} [SM'20] is a Professor with the School of Electronic and Information Engineering, Beijing Jiaotong University. His research interests include cell-free massive MIMO, XL-MIMO, and RIS. He was an Associate Editor for IEEE Transactions on Communications and IEEE Transactions on Wireless Communications.
\vspace{0.1cm}

\textbf{Ziheng Liu} received the B.S. degree from the School of Information and Control Engineering, Qingdao University of Technology, Qingdao, China, in 2023. He is currently pursuing the Ph.D. degree with Beijing Jiaotong University, Beijing, China. His research interests include massive MIMO systems, signal processing, and reinforcement learning.
\vspace{0.1cm}

\textbf{Hongyang Du} received the B.S. degree from Beijing Jiaotong University, Beijing, China, in 2021. He is working toward his Ph.D. degree with the School of Computer Science and Engineering, Energy Research Institute at NTU, Nanyang Technological University, Singapore, under the Interdisciplinary Graduate Program. His research interests include semantic communications, generative artificial intelligence, and communication theory.
\vspace{0.1cm}

\textbf{Zhe Wang} received the B.S. degree from the College of Electronic Information, Qingdao University, Qingdao, China, in 2020. He is currently pursuing the Ph.D. degree with Beijing Jiaotong University, Beijing, China. His research interests include massive MIMO systems, signal processing, and performance analysis of wireless systems.
\vspace{0.1cm}

\textbf{Dusit Niyato} [F'17] is a Professor with the School of Computer Science and Engineering, Nanyang Technological University, Singapore. He received B.Eng. from King Mongkuts Institute of Technology Ladkrabang (KMITL), Thailand, in 1999 and Ph.D. in Electrical and Computer Engineering from the University of Manitoba, Canada, in 2008. His research interests are in the areas of sustainability, edge intelligence, decentralized machine learning, and incentive mechanism design.
\vspace{0.1cm}

\textbf{Mohsen Guizani} [F'09] is a Professor of Machine Learning and the Associate Provost at Mohamed Bin Zayed University of Artificial Intelligence (MBZUAI), Abu Dhabi, UAE. His research interests include applied machine learning, artificial intelligence, Internet of Things, smart city, and cybersecurity. He was listed as a Clarivate Analytics Highly Cited Researcher in Computer Science in 2019, 2020 and 2021.
\vspace{0.1cm}

\textbf{Bo Ai} [F'22] is a professor with the State Key Laboratory of Rail Traffic Control and Safety, Beijing Jiaotong University. His interests include high-power amplifier linearization techniques, radio propagation and channel modeling, global systems for mobile communications for railway systems, and LTE for railway systems.

\bibliographystyle{IEEEtran}
\bibliography{IEEEabrv,ref}

\begin{thebibliography}{10}
\providecommand{\url}[1]{#1}
\csname url@samestyle\endcsname
\providecommand{\newblock}{\relax}
\providecommand{\bibinfo}[2]{#2}
\providecommand{\BIBentrySTDinterwordspacing}{\spaceskip=0pt\relax}
\providecommand{\BIBentryALTinterwordstretchfactor}{4}
\providecommand{\BIBentryALTinterwordspacing}{\spaceskip=\fontdimen2\font plus
\BIBentryALTinterwordstretchfactor\fontdimen3\font minus
  \fontdimen4\font\relax}
\providecommand{\BIBforeignlanguage}[2]{{%
\expandafter\ifx\csname l@#1\endcsname\relax
\typeout{** WARNING: IEEEtran.bst: No hyphenation pattern has been}%
\typeout{** loaded for the language `#1'. Using the pattern for}%
\typeout{** the default language instead.}%
\else
\language=\csname l@#1\endcsname
\fi
#2}}
\providecommand{\BIBdecl}{\relax}
\BIBdecl

\bibitem{[1]}
J.~Zhang, E.~Bj{\"o}rnson, M.~Matthaiou, D.~W.~K. Ng, H.~Yang, and D.~J. Love,
  ``Prospective multiple antenna technologies for beyond {5G},'' \emph{IEEE J.
  Sel. Areas Commun.}, vol.~38, no.~8, pp. 1637--1660, Aug. 2020.

\bibitem{[2]}
M.~Matthaiou, O.~Yurduseven, H.~Q. Ngo, D.~Morales-Jimenez, S.~L. Cotton, and
  V.~F. Fusco, ``The road to {6G}: Ten physical layer challenges for
  communications engineers,'' \emph{IEEE Commun. Mag.}, vol.~59, no.~1, pp.
  64--69, Jan. 2021.

\bibitem{[4]}
H.~Iimori, T.~Takahashi, K.~Ishibashi, G.~T.~F. de~Abreu, D.~Gonz\'{a}lez~G.,
  and O.~Gonsa, ``Joint activity and channel estimation for extra-large {MIMO}
  systems,'' \emph{IEEE Trans. Wireless Commun.}, vol.~21, no.~9, pp.
  7253--7270, 2022.

\bibitem{[3]}
Z.~Wang, J.~Zhang, H.~Du, W.~E.~I. Sha, B.~Ai, D.~Niyato, and M.~Debbah,
  ``Extremely large-scale {MIMO}: Fundamentals, challenges, solutions, and
  future directions,'' \emph{IEEE Wireless Commun.}, pp. 1--9, early access,
  2023.

\bibitem{[17]}
H.~Zhang, N.~Shlezinger, F.~Guidi, D.~Dardari, and Y.~C. Eldar, ``{6G} wireless
  communications: From far-field beam steering to near-field beam focusing,''
  \emph{IEEE Commun. Mag.}, vol.~61, no.~4, pp. 72--77, Apr. 2023.

\bibitem{[9]}
C.~Huang, S.~Hu, G.~C. Alexandropoulos, A.~Zappone, C.~Yuen, R.~Zhang, M.~D.
  Renzo, and M.~Debbah, ``Holographic {MIMO} surfaces for {6G} wireless
  networks: Opportunities, challenges, and trends,'' \emph{IEEE Wireless
  Commun.}, vol.~27, no.~5, pp. 118--125, July 2020.

\bibitem{[19]}
Y.~Liu, J.~Xu, Z.~Wang, X.~Mu, and L.~Hanzo, ``Near-field communications: What
  will be different?'' \emph{arXiv:2303.04003}, 2023.

\bibitem{[10]}
M.~Cui, Z.~Wu, Y.~Lu, X.~Wei, and L.~Dai, ``Near-field {MIMO} communications
  for {6G}: {Fundamentals}, challenges, potentials, and future directions,''
  \emph{IEEE Commun. Mag.}, vol.~61, no.~1, pp. 40--46, 2023.

\bibitem{[5]}
S.~Hwang, H.~Lee, J.~Park, and I.~Lee, ``Decentralized computation offloading
  with cooperative {UAVs}: Multi-agent deep reinforcement learning
  perspective,'' \emph{IEEE Wireless Commun.}, vol.~29, no.~4, pp. 24--31, Aug.
  2022.

\bibitem{[7]}
M.~Rahmani, M.~J. Dehghani, P.~Xiao, M.~Bashar, and M.~Debbah, ``Multi-agent
  reinforcement learning-based pilot assignment for cell-free massive {MIMO}
  systems,'' \emph{IEEE Access}, vol.~10, pp. 120\,492--120\,502, Nov. 2022.

\bibitem{MADDPG}
Q.~Fan, Y.~Zhang, Z.~Wang, J.~Li, P.~Zhu, and D.~Wang, ``{MADDPG}-based power
  allocation algorithm for network-assisted full-duplex cell-free mmwave
  massive {MIMO} systems with {DAC} quantization,'' in \emph{Proc. 2022 WCSP},
  2022, pp. 556--561.

\bibitem{[18]}
H.~Lu and Y.~Zeng, ``Communicating with extremely large-scale array/surface:
  Unified modeling and performance analysis,'' \emph{IEEE Trans. Wireless
  Commun.}, vol.~21, no.~6, pp. 4039--4053, June 2022.

\bibitem{[16]}
Z.~Liu, Z.~Liu, J.~Zhang, H.~Xiao, B.~Ai, and D.~W.~K. Ng, ``Uplink power
  control for extremely large-scale {MIMO} with multi-agent reinforcement
  learning and fuzzy logic,'' in \emph{IEEE INFOCOM 2023 - IEEE Conference on
  Computer Communications Workshops (INFOCOM WKSHPS), Hoboken, NJ, USA}, 2023,
  pp. 1--6.

\bibitem{[12]}
J.~C. Marinello, T.~Abr\~{a}o, A.~Amiri, E.~de~Carvalho, and P.~Popovski,
  ``Antenna selection for improving energy efficiency in {XL-MIMO} systems,''
  \emph{IEEE Trans. Veh. Technol.}, vol.~69, no.~11, pp. 13\,305--13\,318, Nov.
  2020.

\bibitem{[8]}
X.~Chai, H.~Gao, J.~Sun, X.~Su, T.~Lv, and J.~Zeng, ``Reinforcement learning
  based antenna selection in user-centric massive {MIMO},'' in \emph{Proc. IEEE
  VTC2020-Spring}, 2020, pp. 1--6.

\end{thebibliography}
\end{document}